\documentclass[showpacs,preprintnumbers,twocolumn,10pt,aps]{revtex4}
\usepackage{amsmath}
\usepackage{dcolumn}
\usepackage{bm}
\usepackage{graphicx}
\usepackage{amsfonts}
\usepackage{amssymb}

\begin{document}

\title{Entanglement preservation for multilevel systems under non-ideal
pulse control}
\author{Z. Y. Xu$^{1,2}$}
\email{zhenyuxu.wipm@gmail.com}
\author{M. Feng$^{1}$}
\email{mangfeng@wipm.ac.cn}
\affiliation{$^{1}$State Key Laboratory of Magnetic Resonance and Atomic and Molecular
Physics, Wuhan Institute of Physics and Mathematics, Chinese Academy of
Sciences, Wuhan 430071, China\\
$^{2}$Graduate School of the Chinese Academy of Sciences, Beijing 100049,
China}

\begin{abstract}
We investigate how to effectively preserve the entanglement between two
noninteracting multilevel oscillators coupled to a common reservoir under
non-ideal pulse control. A universal characterization using fidelity is
developed for the behavior of the system based on Nakajima-Zwanzig
projection operator technique. Our analysis includes the validity of the
approximation method and the decoherence-suppression by the non-ideal pulse
control. The power of our strategy for protecting entanglement is
numerically tested, showing potential applications for quantum information
processing.
\end{abstract}

\maketitle

Entanglement is a distinctive feature of quantum correlation \cite%
{entanglement} and has played a key role in quantum information processing
\cite{book QIP}. However, due to an unavoidable interaction with the
surrounding environment \cite{book open}, the entanglement among realistic
quantum systems is fragile or may even disappear completely after a finite
interval, known as entanglement sudden death \cite{ESD}. Thus, seeking
entanglement protection among open quantum systems becomes a rewarding but
challenging task in quantum information science. Recently, a variety of
strategies to combat disentanglement have been proposed, including quantum
error correction \cite{QEC}, decoherence free subspaces \cite{DFS}, qubits
embedded in structured reservoirs \cite{structured R}, and vacuum-induced
coherence on the entanglement \cite{interfere}. In addition, the most widely
used methods in experiments are with dynamical control by external fields,
such as quantum feedback control \cite{feedback}, dynamical decoupling \cite%
{DD}, and quantum Zeno effect \cite{zeno}.

However, most of the above mentioned concerns are focused on the well
defined two-level systems, i.e., qubits. For multilevel systems, the
interaction between the environment and systems may cause disentanglement
due to leakage outside of the protected subspace or even the encoded
subspace \cite{nonideal control}. Although we may in principle suppress the
disentanglement by employing ideal bang-bang (BB) control, it could not work
well in real physical systems due to the required arbitrarily strong and
instantaneous pulse constraints \cite{BB}.

In this work, we would like to answer following questions: (i) How to
describe the leakage induced disentanglement of multilevel systems in a
universal form? (ii) Whether and how well can we preserve entanglement of
multilevel systems by non-ideal pulse control? For these purposes, we study
a model of two noninteracting multilevel oscillators resonantly coupled to a
common reservoir under realistic pulse control without the assumption of
idealized zero-width pulses. We will develop a universal description using
fidelity for the leakage induced disentanglement of the open quantum system
based on Nakajima-Zwanzig projection operator technique \cite{book open}. To
our knowledge, it is the first time to present such a universal expression,
which is of no particular dependence on the choice of the entangled states
to be protected. We will present an example to show how well our strategy
can protect entangled states from leakage or decoherence under non-ideal
pulse control. Importantly, due to the relatively free constraints for the
pulses, our strategy can be well met in a variety of experimental situations.

\begin{figure}[!hbp]
\centering
\includegraphics[width=1.2in]{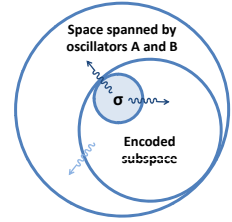}
\caption{(color online). Leakage induced disentanglement of two
noninteracting multilevel oscillators A and B coupled to a common reservoir.
The outer, middle, and inner circles represent a set of subspaces spanned by
combined oscillators A and B, the encoded subspace, and the entangled state $%
\protect\sigma $ to be protected (i.e., protected subspace), respectively.
The wavy arrows represent leakage.}
\label{fig1}
\end{figure}

We first consider two identical multilevel oscillators $A$ and $B$
undergoing longitudinal decay into a common zero-temperature bosonic
reservoir. The resonant frequency $\omega _{0}$ and the dipolar coupling to
the reservoir can be dynamically modulated by external fields with AC-Stark
shifts $\lambda (t)$ and interaction strength $\varepsilon (t)$. The total
Hamiltonian of the composite two-oscillator system plus the reservoir can be
given by $\mathcal{H}=\mathcal{H}_{S}+\mathcal{H}_{R}+\mathcal{H}_{I},$
where (with $\hbar =1$)
\begin{equation}
\mathcal{H}_{S}=\mathbf{[}\omega _{0}+\lambda (t)\mathbf{]}\left(
a_{A}^{\dag }a_{A}+a_{B}^{\dag }a_{B}\right) ,
\end{equation}%
\begin{equation}
\mathcal{H}_{R}=\sum_{l}\omega _{l}b_{l}^{\dagger }b_{l},
\end{equation}

\begin{equation}
\mathcal{H}_{I}=\left( a_{A}^{\dag }+a_{B}^{\dag }\right)
\sum_{l}\varepsilon (t)g_{l}b_{l}+\text{H.c.},
\end{equation}%
are Hamiltonians of the controlled system, the reservoir and their
interaction \cite{DD,book Qoptics}. $a_{j}$ $(a_{j}^{\dag })$ $(j=A,B)$ is
the annihilation (creation) operator of the $j$th oscillator. $\omega _{l}$
and $a_{l}$ $(a_{l}^{\dagger })$ are, respectively, the frequency and the
annihilation (creation) operator of the $l$th mode of the reservoir with the
coupling $g_{l}$ to the oscillators. In the interaction picture, the
Hamiltonian above is rewritten as
\begin{equation}
\mathcal{H}_{I}(t)=S(t)R(t)+S^{\dag }(t)R^{\dag }(t),
\end{equation}%
where $S(t)=\mathcal{M}(t)(a_{A}^{\dag }+a_{B}^{\dag })$, with $\mathcal{M}
(t)=\varepsilon (t)e^{i\int_{0}^{t}d\xi \lambda (\xi )}$ the modulation
function, and $R(t)=\sum_{l}g_{l}b_{l}e^{i(\omega _{0}-\omega _{l})t}.$ The
dynamics of the density matrix $\rho (t)$ of the combined system-reservoir
is governed by the von Neumann equation,
\begin{equation}
\frac{\partial }{\partial t}\rho (t)=-i[\mathcal{H}_{I}(t),\rho (t)]\equiv
\mathcal{L}(t)\rho (t),
\end{equation}%
with $\mathcal{L}(t)$ the Liouville super-operator \cite{book open}. It is
convenient to check that $\left[ \mathcal{H},\mathcal{N}\right] =0$ with the
number operator $\mathcal{N}=a_{A}^{\dag }a_{A}+a_{B}^{\dag
}a_{B}+\sum_{l}b_{l}^{\dagger }b_{l},$ implying $\mathcal{N}$ to be a
conserved quantity. For multilevel oscillators, the system may suffer from
disentanglement due to leakage from the protected subspace (or even from the
encoded subspace) by the interaction with reservoir [shown in Fig. 1]. For
example, we take levels $|0\rangle _{j}$ and $|1\rangle _{j}$ ($j=A,B$) to
span the encoded subspace for qubits, and prepare the initial state of the
system-reservoir in $(\alpha |11\rangle _{AB}+\beta |00\rangle
_{AB})|0\rangle _{R}$, where $|0\rangle _{R}$ denotes the vacuum state of
the reservoir. The system will be losing entanglement or even evolving
outside the encoded subspace e.g., to the state $|20\rangle _{AB}|0\rangle
_{R}$ or $|02\rangle _{AB}|0\rangle _{R}$.

Therefore, our primary aim is to construct a measure for the leakage induced
disentanglement. In the present work, we assume the system to be initially
prepared in an entangled state $\sigma $, which is to be protected. We need
a measure to characterize how far the disentangled state tr$_{R}\{\rho (t)\}$
is away form the protected state $\sigma $ [i.e., we consider the leakage
out of the protected subspace, represented by the dark blue (black) wavy
arrows in Fig. 1]. Such a measure can be $\mathcal{F}(\sigma ,$tr$_{R}\{\rho
(t)\}),$ where
\begin{equation}
\mathcal{F}(x,y)=\left( \text{tr}\sqrt{\sqrt{x}y\sqrt{x}}\right) ^{2},
\end{equation}%
is the fidelity to measure the distance between the density matrices $x$ and
$y$ \cite{book QIP,fidelity}.

Our following aim is to seek a master equation for $\varrho (t)=\mathcal{F}%
(\sigma ,$tr$_{R}\{\rho (t)\})\sigma $, which can be accomplished by
employing the Nakajima-Zwanzig projection operator technique \cite{book open}%
. To eliminate the system-reservoir coherence, we first define the relevant
part of $\rho (t)$ with a super-operator $\mathcal{P}$ \cite{note}:
\begin{equation}
\mathcal{P}\rho (t)=\mathcal{F}(\sigma ,tr_{R}\{\rho (t)\})\sigma \otimes
\rho _{R},
\end{equation}%
where $\rho _{R}$ is supposed to be a stationary state of the reservoir in
thermal equilibrium at zero temperature \cite{book Qoptics}. Since the
initial state of the system is prepared in the state $\sigma ,$ i.e., $%
\mathcal{P}\rho (0)=\sigma \otimes \rho _{R},$ we may get a time-local
homogeneous master equation of $\mathcal{P}\rho (t)$ as
\begin{equation}
\frac{\partial }{\partial t}\mathcal{P}\rho (t)=\mathcal{K}(t)\mathcal{P}%
\rho (t).
\end{equation}%
Here $\mathcal{K}(t)$ describes the time-convolutionless (TCL) generator
\cite{book open}. By restricting Eq. (8) to the second-order expansion of
the system-reservoir coupling, i.e., $\mathcal{K}(t)=\int_{0}^{t}ds\mathcal{%
PL}(t)\mathcal{L}(s)\mathcal{P}$ (the TCL second-order approximation), we
thus obtain the equation for $\varrho (t)$ as,
\begin{equation}
\frac{\partial \varrho (t)}{\partial t}=-\int_{0}^{t}ds\mathcal{F}(\sigma ,%
\text{tr}_{R}\{[\mathcal{H}_{I}(t),[\mathcal{H}_{I}(s),\varrho (t)\otimes
\rho _{R}]]\})\sigma .
\end{equation}%
Inserting $\varrho (t)=\mathcal{F}_{tcl}(t)\sigma $ into Eq. (9) with the
notation $\mathcal{F}_{tcl}(t)=\mathcal{F}(\sigma ,tr_{R}\{\rho (t)\}$, we
obtain
\begin{equation}
\mathcal{F}_{tcl}(t)=\exp \left[ -\int_{0}^{t}d\tau \int_{0}^{\tau }ds%
\mathbf{K}(s,\tau )\right] ,
\end{equation}%
where $\mathbf{K}(s,\tau )$ is the integral kernel with the form
\begin{equation}
\mathbf{K}(s,\tau )=\mathcal{F}[\sigma ,G(s)\Theta (\tau ,\tau -s)+\text{H.c.%
}],
\end{equation}%
$G(t)=$tr$_{R}\{R(t)R^{\dag }(0)\rho _{R}\}=\sum_{l}|g_{l}|^{2}e^{i(\omega
_{0}-\omega _{l})t}=\int d\omega J(\omega )e^{i(\omega _{0}-\omega )t}$ is
the reservoir correlation function with the sum over all transition matrix
elements related to mode $\omega _{(l)}$. $J(\omega )$ is the spectral
density, characterizing the reservoir spectrum \cite{density}. We could
model $J(\omega )$ by several typical spectrum functions, such as the
Lorentzian or from sub-Ohmic to super-Ohmic forms \cite{density}. In what
follows, we consider, as an example, the oscillators interacting resonantly
with a common reservoir with Lorentzian spectral distribution
\begin{equation}
J(\omega )=\frac{1}{2\pi }\frac{\gamma _{0}\Gamma ^{2}}{(\omega _{0}-\omega
)^{2}+\Gamma ^{2}},
\end{equation}%
where $\gamma _{0}$ is the system decay rate in the Markovian limit and $%
\Gamma $ is the spectral width of the coupling \cite{book open}, which has
been widely employed in quantum optics \cite{book Qoptics}. In addition, $%
\Theta (t_{1},t_{2})=[S(t_{1}),S^{\dag }(t_{2})\sigma ]=\mathcal{M}(t_{1})%
\mathcal{M}^{\ast }(t_{2})\bar{\sigma}$, with $\bar{\sigma}=[(a_{A}^{\dag
}+a_{B}^{\dag }),(a_{A}+a_{B})\sigma ]$ fully depending on the choice of the
protected state $\sigma $. Therefore, we have
\begin{equation}
\mathbf{K}(s,\tau )=\frac{\gamma _{0}\Gamma }{2}e^{-\Gamma s}\mathcal{F}%
\left[ \sigma ,\mathcal{M}(\tau )\mathcal{M}^{\ast }(\tau -s)\bar{\sigma}+%
\text{H.c.}\right] ,
\end{equation}%
and particularly, if the protected state is a pure state $\sigma =|\chi
\rangle \left\langle \chi \right\vert $, Eq. (13) reduces to
\begin{equation}
\mathbf{K}(s,\tau )=\eta \gamma _{0}\Gamma e^{-\Gamma s}\varepsilon (\tau
)\varepsilon (\tau -s)\cos \left[ \int_{\tau -s}^{\tau }d\xi \lambda (\xi )%
\right] ,
\end{equation}%
with $\eta =\left\langle \chi \right\vert \bar{\sigma}|\chi \rangle ,$ where
we have assumed $\varepsilon (t)$ to be real. Eq. (13) or Eq. (14) is one of
the main results in this paper. We mention that Eq. (13) [or Eq. (14)] is
universal for any entangled states to be protected. Let us go to some
details below with two examples.

\textit{Example 1}---We consider the oscillators as qubits, i.e., the
protected state within $\{|0\rangle _{A},|1\rangle _{A}\}\otimes \{|0\rangle
_{B},|1\rangle _{B}\}.$ Specifically, we assume the protected states to be
in an extended Werner-like state (EWL) \cite{Wener}
\begin{equation}
\sigma =r|\epsilon \rangle \left\langle \epsilon \right\vert +\frac{1-r}{4}%
\mathbf{1,}
\end{equation}%
where $r\in \lbrack 0,1]$ and $|\epsilon \rangle =|\Phi \rangle =\alpha
|11\rangle _{AB}+\beta |00\rangle _{AB}$ or $|\Psi \rangle =\alpha
|10\rangle _{AB}+\beta |01\rangle _{AB}$. For the sake of conciseness, we
consider two extreme cases: (i) For $r=1$, the EWL state reduces to the
Bell-like pure state $\sigma =|\epsilon \rangle \left\langle \epsilon
\right\vert $. We can get $\eta =2|\alpha |^{2}$ or $\eta =|\alpha +\beta
|^{2}$ when $|\epsilon \rangle =|\Phi \rangle $ and $|\epsilon \rangle
=|\Psi \rangle $, respectively. (ii) In the case of $r=0,$ the EWL state
becomes totally mixed as $\bar{\sigma}=(|\theta \rangle _{AB}\left\langle
11\right\vert -|00\rangle _{AB}\left\langle 00\right\vert )/2$ with $|\theta
\rangle =|11\rangle _{AB}+(|02\rangle _{AB}+|20\rangle _{AB})/\sqrt{2}$. The
lengthy discussion for the case of $0<r<1$ will be left elsewhere.

\textit{Example 2}---We consider the oscillators as qutrits, i.e., the
protected state within the subspace spanned by $\{|0\rangle _{A},|1\rangle
_{A},|2\rangle _{A}\}\otimes \{|0\rangle _{B},|1\rangle _{B},|2\rangle
_{B}\}.$ We obtain $\eta =4|\alpha |^{2}+2|\beta |^{2}$ when the protected
state is given by $\sigma =|\chi \rangle \left\langle \chi \right\vert $
with $|\chi \rangle =\alpha |22\rangle _{AB}+\beta |11\rangle _{AB}+$ $%
\gamma |00\rangle _{AB}.$ It is straightforward to extend the cases to other
entangled states or higher dimensional systems, and the only difference will
be the value of the coefficient $\mathcal{F}$ (or $\eta $).

It is necessary to check whether and to what extent the second-order TCL
approximation can well characterize the dynamics of the two-oscillator
system undergoing decay to a common reservoir. For convenience of
description, we first ignore the external control, i.e., $\lambda (t)=0$ and
$\varepsilon (t)=1$. In order to make a comparison between the exact and TCL
solutions, we consider the state of oscillator-reservoir initially prepared
in $|\Psi (0)\rangle =(\alpha |10\rangle _{AB}+\beta |01\rangle
_{AB})\otimes |0\rangle _{R},$ i.e., at most one excitation of the reservoir
at an arbitrary time: $|\Psi (t)\rangle =C_{10}(t)|10\rangle _{AB}|0\rangle
_{R}+C_{01}(t)|01\rangle _{AB}|0\rangle _{R}+\sum_{l}C_{l}(t)|00\rangle
_{AB}|1_{l}\rangle _{R}$, where $|1_{l}\rangle _{R}=b_{l}^{\dagger
}|0\rangle _{R}$ is the reservoir state with one excitation in the mode $l$.
Only in this case, is there an exact analytical solution to the fidelity by
directly solving the Schr\"{o}dinger equation $i\frac{\partial |\Psi
(t)\rangle }{\partial t}=\mathcal{H}|\Psi (t)\rangle $ \cite{one exciton}.
For simplicity, considering the initial state of the oscillators in $|\Psi
\rangle =(|10\rangle _{AB}+|01\rangle _{AB})/\sqrt{2},$ one immediately has
\begin{equation}
\mathcal{F}_{ext}^{|\Psi \rangle }(t)=\frac{1}{2}|C_{10}(t)+C_{01}(t)|^{2},
\end{equation}%
where $C_{10}(t)=C_{01}(t)=e^{-\Gamma t/2}\left[ \cosh (\kappa t/2)+(\Gamma
/\kappa )\sinh (\kappa t/2)\right] /\sqrt{2}$ with $\kappa =\sqrt{\Gamma
(\Gamma -4\gamma _{0})}$. The corresponding approximate solution reduces to
\begin{equation}
\mathcal{F}_{tcl}^{|\Psi \rangle }(t)=\exp \left[ -2\gamma _{0}\left( t+%
\frac{e^{-\Gamma t}-1}{\Gamma }\right) \right] .
\end{equation}

\begin{figure}[tbp]
\centering
\includegraphics[width=3.4in,bb=41pt 305pt 539pt 518pt]{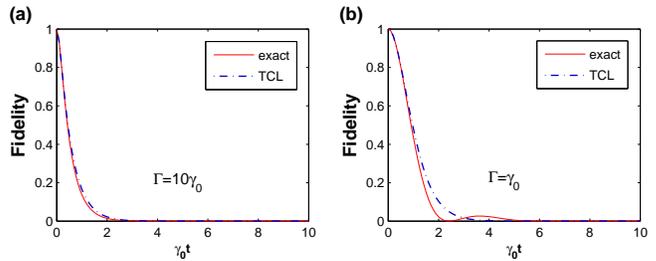}
\caption{(color online). The fidelity for the initial state $(|10\rangle
_{AB}+|01\rangle _{AB})/\protect\sqrt{2}$ as a function of $\protect\gamma %
_{0}t$ with (a) $\Gamma =10\protect\gamma _{0}$ and (b) $\Gamma =\protect%
\gamma _{0}.$ The solid (red) lines and dot-dashed (blue) lines represent
the exact analytical solution and TCL approximate solution, respectively.}
\label{fig2}
\end{figure}

We have plotted Fig. 2 for the validity of TCL approximation, where the TCL
approximation is valid in describing the true dynamics of the system in the
weak coupling regime, i.e., $\Gamma \gg \gamma _{0}$ [See Fig. 2(a)]. But if
$\Gamma$ is comparable or smaller than $\gamma _{0}$, the agreement between
the exact analytical and the approximate solutions occurs only for the
short-time behavior, e.g., $t\leq \Gamma ^{-1}$, as shown in Fig. 2(b).

In what follows, as an illustration, we will concentrate on the entanglement
preservation by some non-ideal pulses by means of Eq. (13) or (14) within
the short-time regime. Unlike idealized BB control, which requires
unrealistic arbitrarily strong and instantaneous control pulses, we are
going to employ non-ideal impulse phase modulation with $\varepsilon (t)=1$
and a periodic rectangular interaction pulse \cite{nonideal control}:
\begin{equation}
\lambda (t)=%
\begin{cases}
\Lambda /\Delta , & for\text{ }nT-\Delta <t<nT, \\
0, & otherwise,%
\end{cases}%
\end{equation}%
where $T$, $\Delta $ and $\Lambda =\int_{nT-\Delta }^{nT}\lambda (t)dt$ are
the period, width and interaction intensity of pulses, which are also the
three main control parameters in the scheme. By numerical treatment, we
seek, using the non-ideal quantum control, the solution to the equation $%
\delta \mathcal{F}_{tcl}(t;T,\Delta ,\Lambda )=0$ with respect to all
control parameters at time $t$ \cite{nonideal control}. As an example, we
employ the pure entangled state to be protected, and the extension to a
general mixed entangled state is straightforward. We assume $\eta =1$
[corresponding to, for example, entangled qubit state $(|11\rangle
_{AB}+|00\rangle _{AB})/\sqrt{2}$ or qutrit state $(|22\rangle
_{AB}+|11\rangle _{AB}+$ $2|00\rangle _{AB})/\sqrt{6}$] and consider the
case with $\Gamma =\gamma _{0}$, which yields the fidelity to be $\mathcal{F}%
_{tcl}(t)=\exp \left[ -\int_{0}^{t}d\tau \int_{0}^{\tau }ds\gamma
_{0}^{2}e^{-\gamma _{0}s}\cos (\int_{\tau -s}^{\tau }d\xi \lambda (\xi ))%
\right] $. This makes it possible to study the entanglement preservation on
the timescale of $\gamma _{0}^{-1}$.

\begin{figure}[tbp]
\centering
\includegraphics[width=3.4in,bb=18pt 203pt 553pt 640pt]{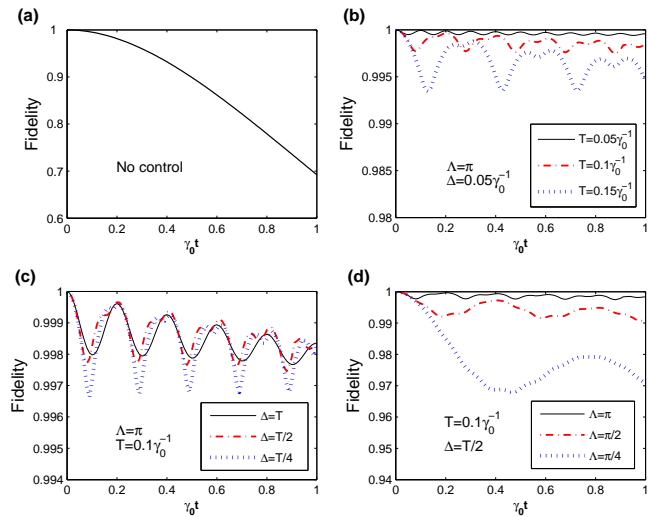}
\caption{(color online). The fidelity for an initial pure state with $%
\protect\eta =1$ [e.g., for $(|11\rangle _{AB}+|00\rangle _{AB})/\protect%
\sqrt{2}$ and $(|22\rangle _{AB}+|11\rangle _{AB}+$ $2|00\rangle _{AB})/%
\protect\sqrt{6}$] as a function of $\protect\gamma _{0}t$ in non-Markovian
regime $\Gamma =\protect\gamma _{0}$, where (a) no external field control is
performed; (b) $\sim $(d) non-ideal impulsive phase modulation of the
coupling with fixed $\Lambda $, $\Delta ;\Lambda $, $T;$ and $T$, $\Delta $,
respectively. The solid (black) lines represent the approximate solutions of
$\protect\delta \mathcal{F}_{tcl}(t;T,\Delta ,\Lambda )=0$ with respect to
variable control parameters $T$, $\Delta $, and $\Lambda $ at an arbitrary
time $t$, respectively.}
\label{fig3}
\end{figure}

To have a comparison, the fidelities of two oscillators undergoing decay
with and without external field control are depicted in Fig. 3. It is clear
that the fidelity drops fast within the timescale of $\gamma _{0}^{-1}$ in
the absence of the pulse control. In contrast, in the presence of pulse
control, despite the pulses with difference in pulse period $T$, pulse width
$\Delta $ and interaction intensity $\Lambda $, the fidelity would be more
or less maintained. We have found that the pulse control works better with $T
$ decreasing, provided the fixed $\Lambda $ and $\Delta $. As shown in Fig.
3(b), when $\Lambda =\pi $, $T=\Delta =0.05\gamma _{0}^{-1}$, the fidelity
nearly remains to be 1. This may be understood as that the control in this
case equals to an addition of a constant frequency $\Lambda /T$ to the
oscillator frequency $\omega _{0}$, and thereby $\Theta $ is oscillating
fast, i.e., the minimum overlap between the reservoir correlation spectral
and the modulation spectral. So the fidelity keeps nearly 1. But for the
fixed values of $\Lambda =\pi $ and $T=0.1\gamma _{0}^{-1}$, as plotted in
Fig. 3(c), the control seems weakly dependent on the pulse width. Fig. 3(d)
demonstrates the dependence of the control effect on the interaction
intensity $\Lambda $: the increase of $\Lambda $ leads to a better control.

The above analysis can be extended to oscillators under other modulations
with realistic parameters. In addition, in future work we will explore the
details of the leakage, which would be helpful for deeply understanding the
physical mechanism behind the dynamics of multilevel systems as well as for
seeking efficient ways to prevent disentanglement of multilevel systems.
Practically, it would be of great interest to test our strategy
experimentally by real physical systems, e.g., using two entangled atoms
confined in optical microcavities \cite{QED}.

In conclusion, based on the Nakajima-Zwanzig projection operator technique,
we have characterized by fidelity the disentanglement of two multilevel
oscillators coupled to a common reservoir. We have developed a universal
expression which well fits the exact solution in weak coupling regime and
strong coupling regime within a short-time period. We have also investigated
the behavior of the oscillators under non-ideal pulse control, which shows
that entanglement could be well protected with high fidelity. We expect that
our strategy would be useful for better understanding dissipative dynamics
of the multilevel open quantum systems and for better operations in quantum
information processing.

This work is supported by the National Natural Science Foundation of China
under Grant No. 10774163.


\begin{thebibliography}{99}
\bibitem{entanglement} R. Horodecki \textit{et al.}, Rev. Mod. Phys. \textbf{%
81}, 865 (2009).

\bibitem{book QIP} M. A. Nielsen and I. L. Chuang, \textit{Quantum
Computation and Quantum Information} (Cambridge University Press, Cambridge,
England, 2000).

\bibitem{book open} H.-P. Breuer and F. Petruccione, \textit{The Theory of
Open Quantum Systems} (Oxford University Press, Oxford, 2007).

\bibitem{ESD} T. Yu and J. H. Eberly, Science \textbf{323}, 598 (2009).

\bibitem{QEC} I. Sainz and G. Bj\"{o}rk, Phys. Rev. A \textbf{77}, 052307
(2008).

\bibitem{DFS} J. Kempe \textit{et al.}, Phys. Rev. A \textbf{63}, 042307
(2001).

\bibitem{structured R} B. Bellomo, R. L. Franco and G. Compagno, Phys. Rev.
Lett. \textbf{99}, 160502 (2007); B. Bellomo \textit{et al.}, Phys. Rev. A
\textbf{78}, 060302(R) (2008).

\bibitem{interfere} S. Das and G. S. Agarwal, e-print arXiv:1004.0564.

\bibitem{feedback} A. R. R. Carvalho and J. J. Hope, Phys. Rev. A \textbf{76}%
, 010301(R) (2007); A. R. R. Carvalho \textit{et al.}, \textit{ibid.}
\textbf{78}, 012334 (2008).

\bibitem{DD} See, e.g., G. Gordon and G. Kurizki, Phys. Rev. Lett. \textbf{97%
}, 110503 (2006); G. Gordon, J. Phys. B \textbf{42}, 223001 (2009), and
references therein.

\bibitem{zeno} S. Maniscalco \textit{et al.}, Phys. Rev. Lett. \textbf{100},
090503 (2008).

\bibitem{nonideal control} L.-A. Wu, G. Kurizki, and P. Brumer, Phys. Rev.
Lett. \textbf{102}, 080405 (2009).

\bibitem{BB} L. Viola and S. Lloyd, Phys. Rev. A \textbf{58}, 2733 (1998).

\bibitem{book Qoptics} M. O. Scully and M. S. Zubairy, \textit{Quantum Optics%
} (Cambridge University Press, New York, 1997); D. F. Walls and G. J.
Milburn, Quantum Optics (Springer-Verlag, Berlin, 2008).

\bibitem{fidelity} R. Jozsa, J. Mod. Opt. \textbf{41}, 2315 (1994).

\bibitem{note} Our evaluation of entanglement preservation is not restricted
to the use of fidelity, but for any measure of the distance between two
density matrices, e.g., trace distance \cite{book QIP}. However, the
definition here provides an intuitive justification: When the protected
state is a pure state $\sigma =|\chi \rangle \left\langle \chi \right\vert$,
we have $\mathcal{P}\rho (t)=\sigma tr_{R}\{\rho (t)\}\sigma \otimes \rho
_{R}$. So $\sigma$ can be taken as a projection operator to project a leaked
state to the protected state, which is in agreement with the statement in
Ref. \cite{nonideal control}.

\bibitem{density} A. J. Leggett \textit{et al.}, Rev. Mod. Phys. \textbf{59}%
, 1 (1987).

\bibitem{Wener} R. F. Werner, Phys. Rev. A \textbf{40}, 4277 (1989).

\bibitem{one exciton} B. M. Garraway, Phys. Rev. A \textbf{55}, 2290 (1997).

\bibitem{QED} K. J. Vahala, Nature (London) \textbf{424}, 839 (2003).
\end{thebibliography}
\end{document}